\documentclass[fleqn,12pt,twoside]{article}

\usepackage{graphicx}

\newcommand{\be}{\begin{eqnarray}}
\newcommand{\ee}{\end{eqnarray}}

\newcommand{\AmS}{{\protect\the\textfont2
  A\kern-.1667em\lower.5ex\hbox{M}\kern-.125emS}}

\title{Building  a ``holographic dual'' to  QCD in the 
AdS$_5$: instantons and confinement  }

\author{Edward Shuryak\\
Department of Physics and  Astronomy,\\
University at Stony Brook, NY 11794, USA}

\begin{document}

\maketitle

\begin{abstract}
Recent attempts to find a ``holographic dual''  to
QCD-like theories included a suggestion by Karsh et al
(below referred to as KKSS)
to incorporate confinement via
 a potential quadratically increasing into the 5-th direction
of the $AdS_5$
 space. We show that
the same conclusion 
follows from completely different line of arguments.  If
  instantons are promoted into the 5d space by identifying
the instanton size $\rho$ at the 5-th coordinate,
the background geometry necessarily should be the
AdS$_5$. As I argued already in 1999,
confinement described via ``dual superconductivity''
leads to a factor $exp(-2\pi\sigma\rho^2)$, where $\sigma$
is a string tension, which is nearly exactly identical to
 that suggested by KKSS. This expression is also well supported
by available
lattice data. At the end of the paper we propose a
IR potential  generalized to the nonzero temperatures.
\end{abstract}
\vskip 2cm

1.{\bf AdS/CFT correspondence} is a conjectured duality
 between the conformal $\cal N$=4
supersymmetric
Yang-Mills theory in d=4 dimensions (to be called CFT for short
below) and the  string theory in AdS$_5\times
O_5$, which passed an impressive number of particular tests
 (see review \cite{adscft}). It  
is clearly the most interesting development at the interface
of  string/field theory. It is
especially fruitful in the large $N$ (number of colors), in
which quantum CFT in a strongly coupled regime is dual to 
classical and weakly coupled bulk supergravity.

 The AdS/CFT
 correspondence made it clear that (contrary to beliefs of many people)
a strong coupling regime does not necessarily imply
confinement. Indeed,
its most direct applications to QCD are related to 
 deconfined high-$T$
phase of QCD, known as   quark-gluon
plasma. It has been recently argued that just above
deconfinement transition it  seems to be
 in a strongly coupled regime, 
because it is a near-perfect liquid as seen from 
analysis of RHIC data
(see e.g. \cite{Shu_liquid}).
 In order to understand confinement
 one has to move from conformal $\cal N$=4
 theory to more QCD-like ones, which still have a gravity dual.
Such approaches, known as 
 a ``top-down'' approach is very interesting, but it is not
discussed in this letter.

 Another direction of current active research, the
 ``down-up'' approach sometimes called
 AdS/QCD,  is a search for a 
``holographically dual'' description of QCD. It is done 
by incorporating various phenomenological 
aspects of QCD, promoted into a
 5-dimensional settings. Such models for QCD spectroscopy
 \cite{5dbag} have started with 
simple infrared cut-off  of the $AdS_5$
space.
Recently Karch et al \cite{KKSS} 
have proposed a more refined effective model, based on the 
conjecture that confinement can be implemented 
by a potential, quadratically growing with the 5-th coordinate into
 the IR. KKSS argued that such behavior 
reproduce correctly the
 behavior of Regge trajectories at high excitations in QCD.

In this letter we point out that the KKSS suggestion is in
fact  (nearly exactly) the same as a conjecture 
 suggested in my paper \cite{Shuryak:1999fe}. The correspondence 
between these two works
 is is especially interesting since they originate in
quite different  
 parts of the QCD dynamics: in \cite{Shuryak:1999fe}
the IR probe is done by instantons
rather  than via hadronic spectroscopy.

 In order to elucidate this correspondence
 we will first show how the instanton dynamics 
can be promoted into the 
5-dimensional space-time, with the 5-th coordinate naturally identified
with the instanton size $\rho$. We will then argue that 
the metric of it must indeed necessarily be
that of the $AdS_5$. After that is done, we will see that
a scalar potential is indeed needed to append purely semiclassical
calculation, and will see that its behavior in infrared
is nearly exactly what KKSS have proposed. 
Not only our line of arguments
is completely different from KKSS one, 
we will also speculate at the end that this potential can arise from
``dual superconductor'' picture of confinement by 't Hooft 
and Mandelstam 
and is thus proportional to the condensate of some 
magnetically charged objects.

2.{\bf Instantons} play a very important role in many aspects of QCD,
including meson and baryon spectroscopy,
for a general review see \cite{Schafer:1996wv}. 
 Instantons of course are also known to play important role
in other gauge theories: see e.g. a review by
Dorey et al \cite{Dorey:2002ik} about $\cal N$=4 theory
and AdS/CFT and
 recent
instanton-based derivation \cite{Nekrasov:2004vw} of 
the celebrated Seiberg-Witten solution of the 
$\cal N$=2 SUSY YM theory.

In QCD interaction between instantons and anti-instantons
cannot be treated analytically\footnote{A simple analytical limits
are possible, e.g.
 at high density, when instanton ensemble is exponentially 
dilute and is a weakly coupled Coulomb plasma \cite{Son:2001jm}, with
light
$\eta'$ as its  ``photon''. In 
 the large $N$
limit  and zero density
the $\eta'$ is still light, but the vacuum is still a strongly
interacting liquid \cite{Sch_largeNc}. In the real $N=3,Nf=2-3$
 QCD the instanton ensemble
is very strongly correlated, as seen e.g. from a very large value
of the $\eta'$ mass.
}. Therefore the
so called ``instanton liquid'' model (ILM) which addresses
many-body aspects of instanton ensemble is usually 
studied numerically. For details and references to papers
which has lead to its development the reader should
consult reviews such as \cite{Schafer:1996wv}.
 It  turned out
to be very successful phenomenologically, predicting
large set of correlation functions in quantitative
agreement with
lattice measurements. Furthermore, predicted dominance of instanton-zero-mode
states for lowest eigenstates Dirac eigenvalues was directly
confirmed by lattice studies: those quark states do dominate
quark propagators st distances relevant for the 
lowest hadronic states. 
In this letter we will not discuss any of these applications
: we will only  touch some building blocks of the ILM
to the extent to show that they can be promoted to AdS space.

The partition function of ILM can be be schematically written as
\be
\label{Z}
 Z &=& \frac{1}{N_+!N_-!}\prod_i^{N_++N_-}\int {d\Omega_i d^4x_i
 d\rho_i \over \rho_i^5}
 \, \exp(-S_{glue}+N_f log det(\bar \rho D\!\!\!\!/)-\Phi).
\ee
where we have identified integration over collective coordinates,
the instanton 4d position $x_i$, size $\rho$ and orientation in
color space $\Omega$. We will discuss different effects
related to the exponent subsequently.

The bosonic action $S_{glue}$ in principle should include
the determinant of the moduli space metric (the overlaps of bosonic
modes), but in practice more important are instanton-antiinstanton
interactions. Those were determined from the so called ``streamline''
method: the specific function would not be important,
all we want to point out
 here is that for an instanton-antiinstanton ($IA$) 
pair their $4N+4N$ collective variables
appear in form of only 
two combinations
\be \label{zd}
z=(1/N)Tr[\Omega_I
 \Omega^+_{A} (\hat R_\mu\tau_\mu)],\,\,\,
d_{IA}^2={(x_I-x_A)^2\over \rho_I \rho_A} \ee
The former is called relative orientation factor, 
the orientation matrices
$\Omega$ live in $SU(N)$ while the $\tau_\mu$ is the usual
4-vector constructed from Pauli matrices: 
 it is nonzero only in the 2*2 corner of the $N*N$ color space.
The 
$\hat R$ is the unit vector in the direction
of 4d inter-particle distance $x_I-x_A$. The second
 combination of position and sizes appears
 due  to conformal invariance of the classical YM theory.
  
The most complicated part of the effective action
is the term which takes into account fermionic exchange
described by 
the determinant of the Dirac operator in the basis 
spanned by the zero modes
\be 
\label{D_zmz}
i  D\!\!\!\!/ &=& \left (
\begin{array}{cc}  0 & T_{IA}\\
                   T_{AI} & 0 
\end{array} \right ).
\ee
The overlap matrix elements $T_{IA}$ are defined by
\be
\label{def_TIA}
T_{IA} &=& \int d^4 x\, \psi_{0,I}^\dagger (x-z_I)iD\!\!\!\!/
\psi_{0,A}(x-z_A),
\ee
where $\psi_{0,I}$ is the fermionic zero mode.
Note that each matrix element has the meaning of an amplitude of a
``jump'' of a quark
from one pseudoparticle to another. 
Furthermore, the determinant of the matrix  (\ref{D_zmz}) is also equal
to a sum of all closed loop diagrams, and thus ILM
 sums all orders in 't Hooft effective $2N_f$-fermion effective
interaction. 

 The matrix element can be written as ($\bar\rho=\sqrt{\rho_I\rho_A}$)
a function of the same two variables (\ref{zd}) as the gauge action
\be
\label{TIA_sum_par}
  \bar \rho T_{IA} = z_{IA} f(d_{IA})\hspace{1cm}
f\approx
  \frac{4 d_{IA}}{\left(2+d_{IA}^2\right)^2}.
\ee
where the expression is an approximate simple parameterization
of a more complicated exact result\footnote{We have shown it to
remind that at large  distances it decays as the distance cube,
corresponding to a fermion exchange in 4d, but $not$ in $AdS_5$. 
}. The reason for that
 is again a conformal
symmetry of zero modes with massless quarks.

3.Promotion of instantons into AdS space is natural for 
$\cal N$=4 theory, see \cite{Dorey:2002ik}. The instantons are
 identified as some point-like objects positions
at distance $\rho$ from the $D_3$ brane. 
Its image at the brane, readily
calculated from the dilaton
bulk-to-brane propagator, gives precisely correct
$G^2_{\mu\nu}(x)$. Furthermore,
remaining 8 supersymmetries near the instanton solution
nicely relate the fermionic zero modes to bosonic ones.
In fact a demonstration by Dorey et al that
in this theory instantons fit so comfortably into AdS/CFT
correspondence
 was one of its early spectacular conformations.

For QCD-like theories, with fundamental (rather than adjoint)
 fermions and no scalars or
supersymmetries, the situation is quite different.
 We will however still argue that one  
can actually lift  an instanton ensemble 
from the brane to bulk in this case as
well, which we will do in two steps.

Step one, same as in $\cal N$=4, is identifying the factor in the
instanton measure  as that in the $AdS_5$ with the metric
\be 
 ds^2={(d\rho^2+dx_\mu^2)\over \rho^2} \hspace{2cm} d^5x \sqrt{g}= {d^4 x d\rho\over \rho^5} 
\ee 
These standard coordinates naturally relate
the $\rho\rightarrow 0$ limit with the ultraviolet (UV)
 and  $\rho\rightarrow \infty$ with the infrared (IR) directions.

Step two requires a good look
at the variable $d$ in (\ref{zd}) for
bosonic and zero-mode interactions between instantons and
antiinstantons:
in fact it is nothing else as
the
 invariant distance between 2 points in  AdS$_5$.

3.{\bf Scale-depending potential}
 $\Phi$ in the partition function (\ref{Z}) contains
all effects that break conformal symmetry of classical YM.
In the   UV it is defined by 
the asymptotic freedom of QCD\footnote{By the way, it is surprising
that this aspect was not included in spectroscopic applications
\cite{5dbag,KKSS}.
}:
\be exp(-\Phi)|_{\rho->0}=(\rho *\Lambda_{QCD})^{(11/3)N-(2/3)N_f}  \ee

Its  IR
behavior is of course non-perturbative and is the main issue discussed
in this work.
 It was the main subject my paper \cite{Shuryak:1999fe},
which argued that in the dual superconductor
model, in which there is a nonzero Higgs VEV of magnetically charged
objects,
one can rely on selfduality of instantons and thus go into
dual ``magnetic'' description. If so, the steps dual to 't Hooft
derivation of instanton measure in Higgs models (such as
the standard model of electroweak interactions) lead to a result 
 that the cutoff is clearly done by
a quadratic potential, $\Phi(\rho)\sim  \rho^2$.

 Moreover, a specific Abelian model for dual superconductor,
identifying QCD strings with Abrikosov vortices, allows to
express the coefficient in terms of the string tension, namely
\be exp(-\Phi)|_{IR}=exp(-2\pi\sigma \rho^2)\ee
It is  the same as proposed by
KKSS, except for numerical factor (which is $\pi/2 $ in their case,
as far as i can tell). 
It was shown in \cite{Shuryak:1999fe} (see Fig.1 borrowed from it)
that this expression very nicely explains the instanton size 
distribution measured on the lattice. (We mention this fact
to stress that we dont see any possibility that the factors of two in
the derivation got confused.)

\begin{figure}[t]
 \includegraphics[width=8cm]{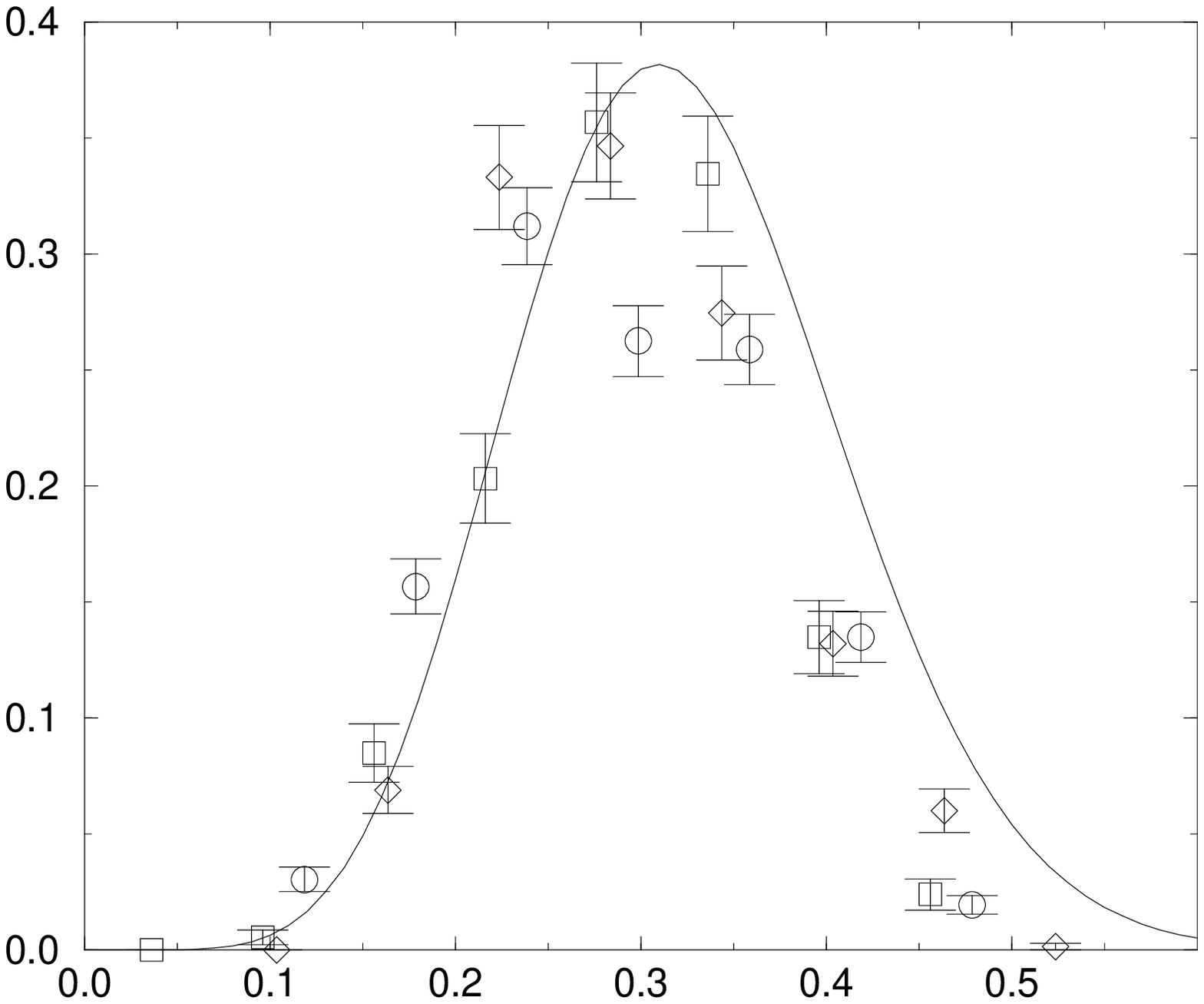}
  \includegraphics[width=8cm]{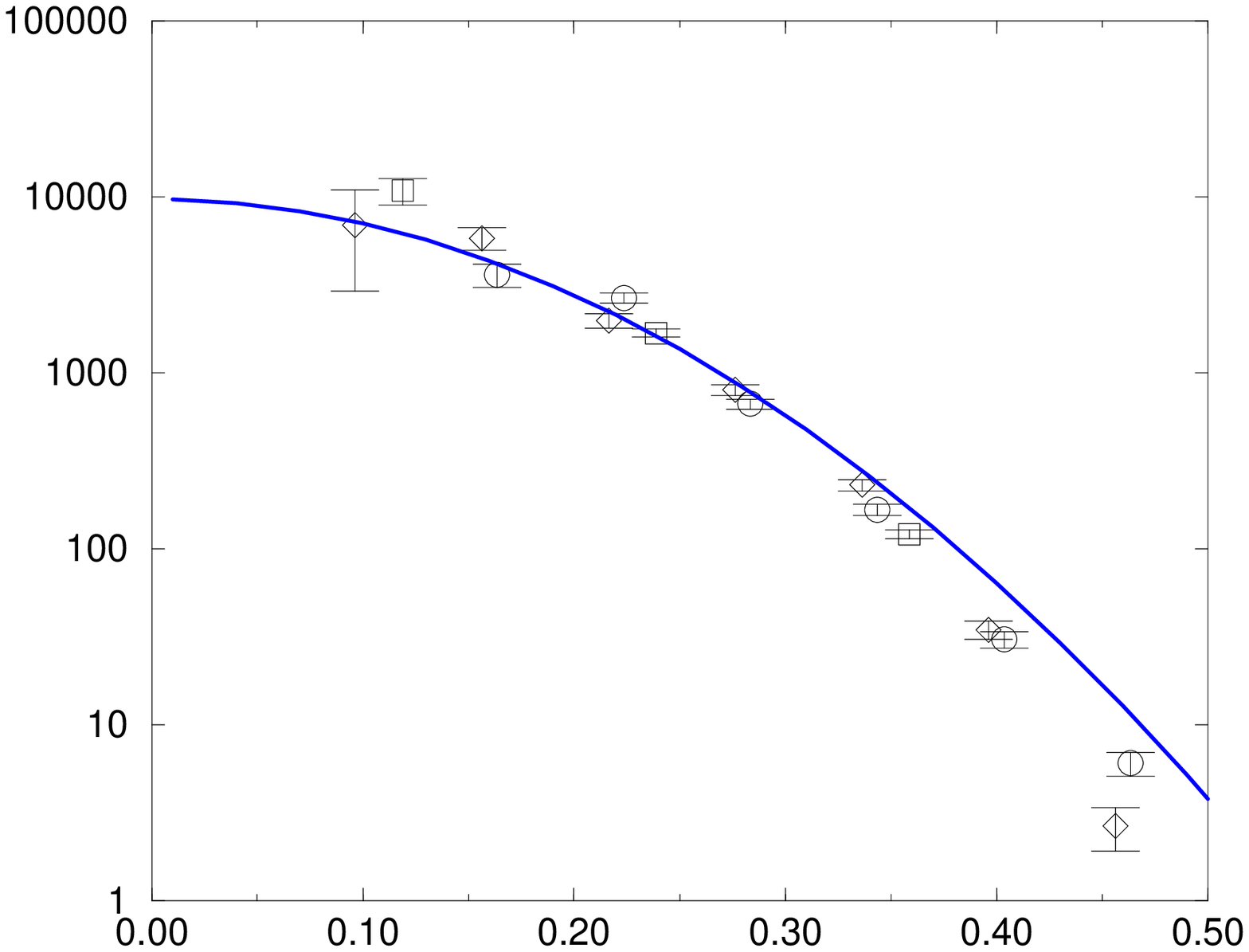}
  \vspace{-.05in}
  \caption{
(a) The instanton density $dn/d\rho d^4x$, [fm$^{-5}$] versus its size
 $\rho$ [fm]. (b) The combination  $\rho^{-6} dn/d\rho d^4z$, in which
 the main one-loop UV behavior drops out for $N=3,N_f=0$, thus
 presumably giving only the IR part of the potential we discuss.
 The points are from the lattice study \protect\cite{anna},
for pure gauge theory, with 
$\beta$=5.85 (diamonds), 6.0 (squares) and 6.1 (circles). (Their
comparison should demonstrate that results are
rather lattice-independent.)
The line corresponds to the 
expression $\sim exp(-2\pi\sigma\rho^2)$, see text.
  }
\end{figure}

Different coefficients of the quadratic term with otherwise identical
parametric dependence is puzzling. One question one might ask
is whether the potential $\Phi$ in the
instanton size distribution is renormalized by 
mutual effective repulsion of 
instantons and antiinstantons\footnote{
In fact in the mean field approach, convergence of the rho integration
was first attributed to it entirely \cite{DP}.}.
However it is easy to see that this cannot  happen
in a  context just described. Indeed, if all the interactions
in an instanton liquid are conformal,
 instantons would be just floating
in the $AdS_5$ and their mutual repulsion would obviously
be related to their density gradient, leading to
homogeneous population of all $AdS_5$ space, as it happens in conformal
theories such as $\cal N$=4. Thus $\Phi$ potential should be
external to a conformal version  of ILM outline above, 
including quantum effects like asymptotic freedom or confinement.

4.{\bf IR cutoff at finite temperatures} is the last issue
we address. If the ``soft cutoff'' is provided by a quadratic
potential proportional to the string tension, as argued above,
this cutoff
 should vanish at the deconfinement as $\sigma(T)\rightarrow 0$
as $T\rightarrow T_c$ and remains zero above it.

In the opposite limit of very high $T$ the instanton suppression
 should be given
by (a perturbatively
calculated) Pisarski-Yaffe \cite{Pisarski:1980md} factor, its main
 part is 
\be exp(\Phi)|{large\,\,T}\sim exp\left[-\frac{1}{3}(2N+N_f)
    (\pi\rho T)^2\right]\ee
Note that a combination of color and flavors are the same
as the perturbative Debye screening
mass.  It is not a coincidence:
as shown in 
  \cite{Shuryak:1982hk}, one can derived it using 
the universal gluon/quark forward  scattering amplitudes on
a (small size) instanton
and the perturbative thermal factors. This leads to natural
generalization of this factor 
\be 
  (N/3+N_f/6) T^2 \rightarrow M_E^2(T)/g^2\ee
where the electric screening mass in the r.h.s.
in not necessarily the lowest order one. It may e.g. include
 in the thermal weight the
 nonzero effective quasiparticle masses and can be independently
determined from lattice simulations. Unlike its perturbative 
counterpart, the screening mass should vanish at and below $T_c$,
where there is no quark-gluon plasma of free charges.
Thus, we have two IR factors so far, one acting below and one above
$T_c$; while both have to vanish $at$ the transition point.

We would now argue that the there must be the third term
in the $O(\rho^2)$ potential, which would be nonzero (and in fact
peaked) at $T_c$: the $magnetic$ screening term. Indeed, 
apart of Bose condensed monopoles there should be also thermally
excited magnetically charged quasiparticles. Their presence
can be monitored via a magnetic screening mass $M_M$,
which is absent in the perturbation
theory but is measured
 the lattice (see e.g.\cite{Nakamura:2003pu}) and seem to have a
 maximum at $T_c$.
 Because our probe - instantons - are selfdual, let me suggest that
the IR cutoff should also respect an electric-magnetic duality. We thus
suggest the IR suppression factor in the form
\be   
exp(-\Phi(\rho,T))|_{IR}=exp[-2\pi\rho^2(\sigma(T)+ {\pi M_E^2(T)\over g^2
  }+{\pi M_M^2(T)\over g_M^2 })]
\ee
where $g_M$ is the magnetic charge. Three terms in the exponent
can all be independently measured on the lattice at any $T$:
we remind that they originate from condensed monopoles, thermal
(non-condensed)
electric and magnetic quasiparticles, respectively.

{\bf Acknowledgments}
The author thanks M.Stephanov for a discussion which
prompted this paper. The work is supported by 
the US-DOE grants DE-FG02-88ER40388
and DE-FG03-97ER4014.

\end{document}